\documentclass[preprint]{aastex}

\slugcomment{Draft version of \today}

\shorttitle{Hot Flux Ropes in the Low Corona}
\shortauthors{A. Nindos et al.}

\begin{document}


\title{How Common are Hot Magnetic Flux Ropes in the Low Solar Corona? A 
Statistical Study of EUV Observations}

\author{A. Nindos and S. Patsourakos}
\affil{Section of Astrogeophysics, Physics Department, University
of Ioannina, Ioannina, Greece}
\email{anindos@uoi.gr}

\author{A. Vourlidas}
\affil{The Johns Hopkins University Applied Physics Laboratory, Laurel MD,
USA}

\and

\author{C. Tagikas}
\affil{Section of Astrogeophysics, Physics Department, University
of Ioannina, Ioannina, Greece}

\begin{abstract}
We use data at 131, 171, and 304 \AA\ from the Atmospheric Imaging
Assembly (AIA) aboard the Solar Dynamics Observatory (SDO) to search
for hot flux ropes in 141 M-class and X-class solar flares that
occurred at solar longitudes equal to or larger than 50\degr. Half of
the flares were associated with coronal mass ejections (CMEs).  The
goal of our survey is to assess the frequency of hot flux ropes in
large flares irrespective of their formation time relative to the
onset of eruptions. The flux ropes were identified in 131 \AA\ images
using morphological criteria and their high temperatures were
confirmed by their absence in the cooler 171 and 304
\AA\ passbands. We found hot  flux ropes in 45 of our events (32\% of
the flares); 11 of them were  associated with confined flares while
the remaining 34 were associated  with eruptive flares. Therefore
almost half (49\%) of the eruptive events  involved a hot flux rope
configuration. The use of supplementary Hinode X-Ray Telescope (XRT) 
data indicates that these percentages should be considered as lower limits 
of the actual rates of occurrence of hot flux ropes in large flares.   
\end{abstract}


\section{Introduction}

Coronal mass ejections (CMEs) are large-scale expulsions of coronal
plasma and magnetic field entrained therein into the heliosphere.
Several models of CME initiation have been developed (e.g. Chen 2011,
and references therein). All models agree that CMEs may result from a
catastrophic loss of mechanical equilibrium of plasma confined by the
coronal magnetic field. However, there is no consensus about the
pre-eruptive magnetic configuration; the CME models can be divided
into two groups depending on the state of the coronal magnetic field
prior to the eruption.  The models of the first group assume that a
magnetic flux rope (i.e. a coherent magnetic structure of magnetic
field lines that collectively wind about a central, axial field line)
exists prior to the eruption (e.g. Forbes and Isenberg 1991; Forbes
and Priest 1995; Gibson and Low 1998; Titov and D\'{e}moulin 1999;
Roussev et al. 2003; Amari et al. 2004; 2005; Fan and Gibson 2004;
2007; Archontis and T\"{o}r\"{o}k 2008; Fan 2010; Archontis and Hood
2012). The models of the second group rely on the existence of sheared
magnetic arcades (i.e. an arrangement of loops with planes deviating
significantly from the local normal to the polarity inversion line)
that become unstable  and erupt once some critical state is reached in
the corona (e.g. Linker  and Miki\'{c} 1995; Amari et al. 1999; 2000;
Antiochos et al. 1999;  Manchester 2003; Roussev et al. 2004; MacNeice
et al.  2004; Jacobs et al. 2006; van der Holst 2007; Lynch et
al. 2008; Archontis and Hood  2008; Karpen et al. 2012).

While the question on which pre-eruptive magnetic configuration 
leads to CMEs is open, all models and simulations agree that the
erupting structure is  a flux rope. There is no physical
mechanism that can produce a large-scale eruption from the corona
without ejecting a flux rope, except under very specific magnetic configurations (Jacobs et al. 2009). In the first group of models the flux
rope is an essential ingredient of the pre-eruptive configuration
while in the second group of models the flux rope is formed once the
CME is underway, i.e., on-the-fly. 

Indirect evidence for the existence of pre-eruptive flux ropes in
active regions (ARs) located close to disk center comes from
non-linear  force-free field (NLFFF) extrapolations that use data from
photospheric  vector magnetograms as boundary condition (e.g. Yan et
al. 2001; Bleybel  et al. 2002; Canou et al. 2009; Guo et al. 2010;
2012; Sun et al. 2012; Nindos et al. 2012; Jiang \& Feng 2013; Cheng
et al. 2014b). The temporal resolution  of NLFFF extrapolations is
determined by the cadence of the available vector magnetograms which,
is currently 12 minutes, at best. Therefore, the extrapolation products 
are unable to capture the rapidly-changing magnetic fields that erupt. 
Furthermore,  there is a non-vanishing Lorentz force that makes the magnetic 
field non-force-free, once the eruption gets underway.

Another piece of indirect evidence for pre-existing flux ropes are
soft X-ray (SXR) and EUV observations of  S- or reverse S-shaped
(sigmoidal) regions.  Sigmoidal ARs have a high likelihood of
producing an eruption (e.g. Canfield et al. 1999). If a sigmoidal
source of SXR or EUV emission, which follows the magnetic field
lines, crosses the polarity inversion line in the inverse direction of
what a potential arcade would do then it is considered as disk
signature of a flux rope viewed from above. This, however, is
conclusive only if the sigmoidal source survives an eruption
(e.g. Gibson \& Fan 2006; Green \& Kliem 2009; Green et al. 2011),
because only then S-shaped field lines in a sheared arcade (Antiochos et
al. 1994) can be excluded.

{\it Yohkoh} (Ogawara et al. 1991) Soft X-ray Telescope (SXT;
Tsuneta et al. 1991) data (Ohyama \& Shibata 1998; 2008; Nitta \&
Akiyama 1999; Kim et al. 2004; 2005a,b) and {\it Hinode} (Kosugi et
al. 2007) X-Ray Telescope (XRT; Golub et al. 2007) data (Savage et
al. 2010; Landi et al. 2013; Lee et al. 2015)  have revealed the
existence of hot ejecta during  CMEs. The ejecta formed coherent hot
structures but in most of these publications were not identified as
magnetic flux ropes.

The availability of high sensitivity data recorded with unprecedented
spatial and temporal resolution in hot, flare-like EUV wavelengths by
the Atmospheric Imaging Assembly (AIA; Lemen et al 2012) aboard the
Solar  Dynamics Observatory (SDO; Pesnell et al. 2012) confirmed the
existence of coherent structures identified as hot flux ropes (Reeves
\& Golub 2011; Cheng et al. 2011, 2012, 2013, 2014a,b,c; Zhang et
al. 2012; Li \& Zhang 2013; Patsourakos et al. 2013; Joshi et
al. 2014; Song et al. 2014). The heating of the flux rope may be a
consequence of magnetic reconnection. The standard solar eruption
model predicts that the reconnected magnetic flux under the CME is
channeled into both the flare loops and the erupting flux rope. 

Several case studies of hot flux ropes associated with impulsive CMEs
have been published. The main focus of these publications was the
question of whether the flux rope forms before or during the
eruption. Most studies were in favor of pre-existing flux ropes (Zhang
et al. 2012; Patsourakos et al. 2013; Li \& Zhang 2013;  Cheng et
al. 2013; 2014a; 2014b; 2014c) although cases of hot flux ropes formed
on-the-fly during the eruption (Cheng et al. 2011; Song et al.  2014)
have been reported as well as cases of hot flux ropes involved in
confined eruptions (Patsourakos et al 2013; Cheng et al. 2014b; Joshi
et al. 2014).  All these events originated in active regions close to
the solar limb.

These works were exclusively single-event studies, leaving an unclear
picture on the role/importance of hot flux ropes in CMEs. To remedy this
situation, we undertake the present  study to search for hot flux rope
events using an extensive dataset of large flares that occurred away
from disk center. To keep this paper focused, we do not address the question on
whether the flux ropes formed before or during the eruptions.
Instead, we address the more fundamental question of how common
hot flux ropes are in the low corona. Our aim is to place past and future case
studies in a broader context. Our paper is organized as follows. The
observations and data analysis are presented in Section 2. Our
classification scheme is introduced in Section 3. In Section 4 we
discuss the events  with hot flux ropes and in Section 5 the events
without hot flux rope morphology. In Section 6 we compare the AIA data
with data obtained from the Hinode XRT. We conclude in Section 7.

\section{Observations and Data Analysis}

We have compiled a catalog of M-class and X-class flares that occurred
from early November 2010 until the end of March 2014 at longitudes
equal to or larger than 50\degr\ from central meridian. We selected
events  away from disk center because the identification of flux ropes
becomes easier due to projection effects. We set an M-class threshold
because the higher energy release is more likely to produce hot flux
ropes. We do not preclude the existence of flux ropes in other flare
types.  Our catalog consists of 141 events. We analyzed each event
using EUV images of the low corona from the SDO/AIA. The field of
view, pixel size, and cadence of the AIA images are 1.3$R_{\odot}$,
0.6\arcsec, and 12 s, respectively. To reduce data volume,  we used
AIA images with a  cadence of 1 minute. Inspection of full cadence
movies of selected events showed that the one minute cadence was
sufficient to resolve the various dynamics.
  
We used AIA images in narrowband channels centered at
131, 171, and 304 \AA. In the remainder of the paper we will refer to
any given channel by simply supplying the wavelength of peak response:
for example, 131 \AA\ channel will be referred to as 131. The 131
emission arises from two dominant ions  formed at
 different temperatures: Fe VIII at $4 \times 10^5$ K and Fe
XXI at $10^7$ K (Lemen et al. 2012). The signal in 131 is dominated by
multi-million plasma only during flares. The dominant ions
in the 171 and 304 channels are Fe IX and He II, respectively. Their
peak formation temperatures are $6 \times 10^5$ K and $5 \times 10^4$
K, respectively (Lemen et al. 2012). 

In addition to the AIA data we used white-light coronagraph data from
the Large Angle Spectroscopic Coronagraph (LASCO; Brueckner et al
1995) C2 coronagraph on board the Solar and Heliospheric Observatory
(SOHO; Domingo et al 1995) mission. The LASCO C2 field of view covers
the range of 2.2-6$R_{\odot}$ with a pixel size of 12\arcsec, and
nominal cadence of 12 minutes during the period of our observations. For the overall temporal evolution of
the flares, we used GOES X-ray total flux measurements.

Our first analysis step was to search for signatures of hot flux ropes
in 131 movies. The observed morphology of  flux ropes depends on its
inherent twist and viewing angle but their identification 
is often challenging  because: (1) the hot flux rope emission at
131 \AA\  may be weak, (2) more often than not, flux ropes are weakly
twisted, and (3) other structures (e.g. loops), and saturation from
the flare may contaminate the line of sight (LOS). Based on our experience, we  settled in the following set of criteria for flux rope identification in the 131 images: 

\begin{enumerate}

\item Flux ropes seen edge-on. They correspond to round blobs or ring-like
structures. They appear to overlay structures with a variety
of morphologies: $\Lambda$-shaped loop-like structures, cusp-like
structures or thin elongated emissions presumably associated with current
sheets. Examples of flux ropes seen edge-on appear in Figure~1.

\item Flux ropes seen face-on. The clearest cases show tangled threads
of emission that appear to wind around an axis. Structures that show a twisted/writhed shape could also be interpreted 
as flux ropes seen face-on. Examples of flux ropes seen face-on appear in 
Figure~2.

\item Flux ropes viewed from intermediate angles. They show intermediate
morphologies between the morphologies of flux ropes seen edge-on and face-on. 
Examples are presented in Figure~3.

\item The candidate flux rope should appear in at least two successive 131 
\AA\ images.

\end{enumerate}

Because we are searching for hot flux ropes, the candidate structures
in the 131 \AA\ data must be hot. An obvious approach would be  the
calculation of the differential  emission measure of the structures and the subsequent computation of emission measure maps in
different temperature intervals. We decided against such approach at this time due to the large size of our database and the added complexity of the DEM calculation procedures.
Instead, we adopt a simple, but straightforward approach, to check for the simultaneous appearance of
the candidate flux ropes in 171 and 304 movies. If the candidates do
not appear in these colder channels then the 131 emission must arise
from Fe XXI and hence from 10 MK plasma. We note that this was the case
for all candidate hot flux ropes that we identified in 131 data.

We tried especially hard to avoid over-interpretation of the data. Several
features have the potential to be misinterpreted as flux ropes. These
include cusps, ascending post-CME loops (especially those
seen edge-on) and small-scale emission depletions that are not
surrounded by clear twisted/writhed structures or ring-like
structures, but result from the projection of loops of different
shapes and orientations on  the plane of the sky.  

Finally we used movies of LASCO C2 data to determine whether the
events in our database were associated with CMEs. The association was
checked for all events of our database, irrespective of being hot flux
ropes or not. The CMEs  were also classified as flux rope CMEs
(FR-CMEs) or not using the criteria defined by Vourlidas et
al. (2013). 

\section{Classification of Events}

Our database consists of 141 events shown in Table 1. The
first six  columns  give (from left to right) the event
number, the date of the  event, the flare start and peak times, the
flare location in heliographic  coordinates, the NOAA number of the
host AR, and the GOES classification of the  flare. The seventh column contains the information on the associated CME, if there was one. The eighth column provides
our classification of the events into five subcategories.

Overall, 45 flares were associated with hot flux ropes and 96 flares were not. This simple division, however, does not provide much insight into the problem. We refined our classification to reflect the wealth of observed morphologies and eruption paths
present in the data. According to this refined scheme, the events were classified in the following groups (Table 1, column 8):

\begin{enumerate}

\item Confined flare events with hot flux ropes (CFR in Table 1). No evidence 
of eruption but detection of a hot flux rope in the AIA images.

\item Eruptions with hot flux ropes (EFR in Table 1). Evidence for the 
eruptive nature of the events was provided by
post-flare/post-CME loops, cusps, or overall opening of the AR's configuration.

\item Prominence eruption events without hot flux ropes (PE in Table 1).

\item Eruptions without hot flux ropes or prominences (PFL in Table 1).

\item Confined flare events without hot flux ropes (CFL in Table 1). There was  
no evidence of an eruption or a hot flux rope. 

\end{enumerate}

\section{Events With Hot Flux Ropes}

\subsection{Morphology of Hot Flux Ropes}

We identified 45 events with hot flux rope morphology; in 34 cases
there was an eruption (EFR events of Table 1) and in 11 cases the hot
flux rope did not erupt (CFR events of Table 1). Single snapshots of
various hot flux ropes in 131 \AA\ are presented in Figures~1-3.  In Figure~4 we present the evolution of a confined hot flux rope
event  while in Figures 5-7 we present the evolution of three
eruptive hot flux rope events.

The hot flux ropes can be divided  into three categories according to their LOS orientation (see
Section 2):  (1) hot flux ropes seen edge-on (Figures~1, 4, and 7),  (2) hot flux
ropes seen face-on (Figures~2 and 6), and (3) hot flux ropes viewed from  
intermediate angles (Figures~3 and 5).

Our database contains 20 cases of hot flux ropes seen edge-on, 9 cases
of hot flux ropes seen face-on and 16 intermediate cases. Most of the
previous analyses of hot flux rope events involve flux ropes seen
edge-on. Reports on hot flux ropes seen face-on is rather
limited (e.g. Li \& Zhang 2013, Cheng et al. 2014c, Joshi et
al. 2014).  The larger proportion of flux ropes seen edge-on can be
explained as follows. The LOS integration along a structure seen
edge-on is longer than for a face-on case. Flux ropes are
expected to run approximately along the neutral line of their host ARs
which generally lies along the East-West direction. Therefore, as the
AR approaches the limb, the neutral  line becomes more and more
aligned with a direction parallel to LOS, and consequently its
associated flux rope should be seen edge-on. 

\subsection{Confined Events with Hot Flux Ropes}

The confined nature of the CFR events was judged from the EUV data and 
confirmed from LASCO observations. LASCO data were available for all 11 CFR 
events and none of them showed evidence for CMEs.

Snapshots of confined events with hot flux ropes are presented in
Figure~1 (panels a-d). All of them appear as round blobs of hot plasma
which implies that they are seen edge-on. In the event of panel (a),
there is thin elongated emission resembling a current sheet just
underneath the flux rope. The situation in the event of panels (b),
(c), and  (d) is less clear in terms of the appearance of
current-sheet-like or cusp-like structures, but nevertheless stretched
loops that connect the flux rope with the lower atmosphere can be
seen.

The evolution of a characteristic CFR event (event 86) is presented in 
Figure 4 (see also movie1.mp4). The event took place on 2013 January 5 in 
AR11652 and  was associated with an M1.7-class flare. Figure 4(a) was taken
just before the flare onset. The image shows loops of different
scales; from  low-lying loops to large-scale loops that correspond to
the background  magnetic field. At about 09:29 in 131 \AA\ a
semi-circular bright loop-like  feature appears very close to the
limb. At 09:30 an initially round blob evolves clear striations by 9:31 (Figure 4b,c).  This morphology corresponds to a flux rope seen
edge-on. 
The flux rope appears to sit at the tip of a current-sheet-like
structure (Figure 4b,c). During the next two minutes the flux rope moves
outward increasing in size. According
to the standard flare model, magnetic reconnection induced in the
current sheet converts the stretched surrounding field into new
poloidal flux of the flux rope. In addition to its ascending motion,
it is possible that the flux rope exhibits rotation as well. The
rotation changes the orientation of the flux rope with respect to the
line of sight and this may explain why its morphology has changed in
Figure~4d. We see an elliptical structure with S-shaped threads inside it instead of a round blob. After 09:33  the flux rope
stops rising. The associated movie gives the impression that the
overlying field acts as an obstacle that inhibits the ascending motion
of the flux rope. 

The flux rope is visible only in the 131 passband (see the movie,
Figure 4f, 4h and the  131/171 and 131/304 composites in panels g and
i).  Therefore   the flux rope temperature is about 10 MK, i.e. it is
a {\it hot} flux rope. Furthermore, the 131/171 composite image shows
that the hot flux rope is enclosed in an area of weak 171
\AA\  emission.  Several authors have reported (e.g. Cheng et al.
2011, 2013;  Zhang et al. 2012) that hot flux ropes are sometimes
enclosed in a dark cavity or bubble, observed in cooler emissions,
e.g., in the 171 channel. The hot flux rope may not fully occupy the
cavity, at least  during the initial stages of the EUV cavity
formation, in both confined and  eruptive events (Kliem et al. 2014).

\subsection{Eruptive Events with Hot Flux Ropes}

The evidence for eruption in the AIA data is based on the appearance
of post-CME  loops, cusps or major opening of the AR configuration.
Snapshots of eruptive events with hot flux ropes are presented in
Figures 1 (panels e-f), 2, and 3. They show diverse morphologies which
reflect the different orientation of the flux ropes with respect to
the LOS (see Section 2).

An example of an eruptive hot flux rope event (event 115)  is
presented in Figure 5 (see also movie2.mp4).  The event took place on
2013 November 21 in AR11895 and was associated  with an M1.2-class
flare. The pre-flare configuration is shown in Figure 5(a). According
to the 131 \AA\ movie, from about 10:52  loops start rising while some
of the low-lying loops become brighter.  After 10:55 interaction
between rising loops yields an elliptical blob of emission that pushes
the overlying loops and stretches the whole rising magnetic
configuration. As a result, a cusp is formed underneath the rising
elliptical blob (Figure 5b). We interpret the rising blob as a hot
flux rope seen almost edge-on. In the next few minutes the hot flux
rope grows  (Figure 5c) presumably by reconnection that  feeds it with
new poloidal flux (e.g. Lin \& van Ballegooijen 2002)  and its
morphology evolves from an almost elliptical blob to   a
twisted/writhed structure (panels (c)-(d)). The change in morphology
indicates that, in addition to its growth, the hot flux rope may
exhibit rotation. The flux rope could  be tracked until about 11:10
(Figure 5(e)). Its detection was not possible after that time due
partly to its proximity to the edge of the AIA field of view and
partly  to the shorter exposure times due to the flare.

As in the case of Figure 4, we are confident that the flux rope is hot
because it appears only in the 131 \AA\ data and not in the 171/304
\AA\ data (see the movie, and panels (f)-(i) of Figure 5). The
opening of the AR's magnetic configuration shows clearly from about
11:01 to about 11:07 in the 171 \AA\ movie  as a faint large-scale
loop-like structure rises and eventually opens up. 

An example of the evolution of an eruptive event with a hot flux rope seen 
face-on (event 25) is presented in Figure 6 (see also
movie3.mp4). The event occurred on 2011 September 22 in AR11302 and
was associated with an X1.4-class flare. From the beginning of the
movie the preflare loops (see Figure 6a) rise and the core of the AR
brightens.  The first evidence of the hot flux rope appears around
10:24 as a structure consisting of thin tangled threads
(Figure~6b). For clarity, we delineate its outer edge with a dotted
curve (Figure~6c). Figures~6d-e follow the rise of the flux rope (see also
movie3.mp4). The movie and panels (c), (f)-(i) of Figure 6 indicate
that the flux rope was hot because it was visible only in the 131
\AA\ data.

Our database contains 7 erupting hot flux rope events  associated with
prominence material. One example is presented in Figure~7 (see also
movie4.mp4). The event occurred on 2014 February 9  and was associated 
with an M1.0-class flare (event 130).  In the movie the upward motion of 
material starts at about 15:11 and shows better in 304. From about 15:18 a
twisted structure appears in all three passbands which gradually grows
(Figures 7a-b). At the same interval, the movie shows more
prominence material ejecting in 304 \AA. As time passes the 304
\AA\ prominence  forms  a large twisted structure which appears much
smaller in 131 and 171 passbands. In 131 \AA\ the configuration of the
twisted structure gradually evolves (Figure 7c) to form a hot flux
rope. It appears first at about 15:38, then grows and eventually
leaves the instrument's field  of view. In Figure 7(d)-(e) we mark the
candidate flux rope structure. Its appearance
resembles a flux rope seen edge-on while its spatial scale is
different (i.e. smaller) than the spatial scale of the ejected
material at both 171 and 131 passbands (Figure 7, panels (d)-(i)).
 
Overall the morphologies in event 130 imply that the prominence
eruption contained primarily cool and warm material which was
intermingled with very hot material that showed a flux rope
configuration. The hot flux rope was located inside the upper part of the
large-scale twisted structure that was  formed by the cool ejected
material (Figure 9i). This morphology is similar to the event
presented in Zhang et al. (2012).

There are basically two patterns in the 7 events with both eruptive
hot flux rope and prominence material. In four of them (two of them
are the events of Figures 7 and 3a) the ejected prominence material and
the eruptive hot flux rope intermingled whereas in three of them (one of them
is the event of Figure~3f) the ejected prominence material was segregated
from the hot flux rope that erupts. 

\section{Events Without Hot Flux Ropes}

There are 19 prominence eruption events without signatures of hot flux
ropes in our database (PEs in Table 1). A typical example (event 33 of
Table 1) is presented in Figure 8. In all PE events the erupting prominence material was detected  in all three AIA passbands. At 304 \AA\ we detect cool chromospheric plasma while at 171 and
131 passbands we detect warmer plasmas that may come either from the
prominence-corona transition region (the warm plasma between the
building blocks of the prominence and the adjacent hot corona,
e.g. Luna et al. 2012) or from the mild heating to temperatures in
excess of $2 \times 10^5$ K that erupting prominences sometimes
experience (e.g. Landi et al. 2010).

LASCO observations were available for 18 of the 19 PE events (there was
a LASCO data gap for event 57). An inspection of the LASCO movies showed 
that 17 out of the 18 PE events were associated with CMEs. This is
consistent with the well-known association of prominence/filament
eruptions  to CMEs (i.e., Gopalswamy et al. 2003 reported that 72\%
of the filament eruptions they studied were associated to CMEs). We note 
that 5  of the 17 CMEs were flux rope CMEs.

We want to make clear that the lack of \textit{hot\/} flux rope
signatures for these 19 PE events does not preclude in any way the
existence of a flux rope. Five of the events were associated with a
flux rope CME after all.  We merely argue there was no trace of {\it hot}
flux ropes, according to our criteria in Section 2.

The database contains 24 eruptive events without the presence of a hot
flux rope or a prominence  (PFLs in Table 1). Figure~9 presents an
example of a PFL event (event 72) associated with an M1.0-class flare
on 2012 August 17 in AR11548. The top row shows the evolution of the
flare in 131 \AA; in panel (a) we present a pre-flare image while the
image of panel (b) corresponds to one minute before the flare peak. In
panel (b) a cusp-shaped morphology appears as well as bright loops
underneath the cusp. These features are in agreement with the standard
model of eruptive events. In panel (c), taken about 7 minutes after
the flare peak, the cusp is still present although the reverse Y
morphology of panel (b) is not so prominent, Just above the tip of the
cusp, a faint hollow elliptical feature has been marked by an
arrow. That feature could be misinterpreted as hot flux rope. However,
we could not follow it with certainty in subsequent images and its
appearance may not be associated with a unique physical entity;
instead,  its appearance may come from the alignment of the emission
from the tip of the cusp with faint background emission along the line
of sight. Panel (d) corresponds to the decay phase of the flare; the
cusp morphology is still present, however its height  has decreased.
The cusp does not show in the 171 \AA\ data (see middle row of Figure
9). It is possible that its emission was either obscured by the
bright background emission or that it did not have the right temperature to
show in 171 data. However, the cusp appears at 304 \AA\ (bottom row of 
Figure 9) several minutes after the flare  maximum (panel i) and it is probably Si XI 303.4 \AA\ line emission from 1.8 MK plasma (e.g., Figure~4 in Stenborg et al. 2008). 

Inspection of LASCO movies shows that 20 out of the 24 PFL
events were associated with CMEs, thus confirming the eruptive nature 
of those events.

We classified as PFLs all eruptive events that did not
show clear prominence eruptions and for which we were not confident
about the presence of hot flux ropes (see the discussion in Section 2
about the care we took to avoid over-interpetation of the data). But 10 out
of the 20 CMEs associated with PFL events were flux rope
CMEs. Therefore, some fraction (at least) of the PFL events must 
had flux rope  configuration, but it did not conform to our
identification criteria.

The database contains 53 events of confined flares with no signatures
of hot flux ropes  (CFLs in Table 1). The confined nature of the
flares was derived from the study of the AIA data (i.e., lack of dimmings
or EUV waves) and LASCO C2 observations. LASCO data were available for
50 of the events (due to a data gap, there were no LASCO observations
for events 58, 59, and 60). None of them showed evidence for CMEs
in agreement with the AIA data.

In Figure 10 we show a characteristic example of a CFL event (event 50)
that occurred on 2012 February 6 in AR11410. We show  only the 131
passband, because the evolution of the event was similar in the 171
and 304 passbands. The event was a gradual M1.0-class flare.  Panel
(a) shows the AR configuration at the time of the flare start, where
different sets of sheared loops appear. In panel (b), at the rise
phase of the flare, some of the loops appear brighter and their
configuration has been simplified (at least partly). However, a new
set of tangled loops has appeared southwest of the initial
configuration.  The image of panel (c) was taken one minute after the
flare maximum while the image of panel (d) corresponds to the decay
phase of the  flare. Figure 1 indicates that the flare was confined
because no trace of rising arcades, cusps or major opening up of the
of AR's configuration were detected. 

\section{Comparison with Hinode XRT data}

The existence or absence of hot flux ropes in 131-\AA\ data was
checked against data obtained with the XRT on the Hinode mission. 40
events of our catalog were also observed by the XRT.  These flares
were observed in various combinations of thin, medium-thickness,  and
thick XRT filters with cadence ranging from about 30 s to about 2
minutes. The field of view was either $384\arcsec \times 384\arcsec$
(pixel size of about 1\arcsec) or $512\arcsec \times 512\arcsec$
(pixel size of about 2\arcsec) or  $1536\arcsec \times 1024\arcsec$
(pixel size of about 4\arcsec). The thin filters have peak temperature
responses from about 8 MK to about 10 MK while the medium-thickness
and thick  filters have peak temperature responses at about 13
MK. However, all XRT filters are able to detect a wide range of
temperatures (the thin filters above 2 MK) because of their broad
responses (see Narukage et al. 2011, for details on the XRT filter
response functions). Therefore, checking the broadband thin-filter XRT
data may be roughly equivalent to checking data from narrowband AIA
channels centered at 211, 335, and 94 \AA\  whose dominant ions are
formed at 2.0, 2.5, and 6.3 MK, respectively (Lemen et al. 2012).

In Table 2 we present the statistics of the  events observed
by both AIA and XRT. 14 of the 40 flares (35\%)  contained hot flux
ropes in the AIA data (4 CFRs and 10 EFRs). Therefore, the sample of
events observed by both instruments represents a fairly typical subset
of the whole catalog (32\% of the flares of the whole AIA  database
contained hot flux rope morphologies;  see Table 3 and Section 7). We
searched for signatures of flux ropes in the XRT data using the same
morphological criteria that we used for the detection of flux ropes in
131-\AA\ AIA data. 

We found that 43\% (17/40) of the XRT flares showed flux rope
morphologies.  All of the XRT flux ropes were observed with the
thin filters,  while two events were also observed with
the medium-thickess Al filter and one event was also observed with the
thick Be and thick Al filters.  11 out of the 14 events with 131-\AA\ hot
flux ropes ($\sim$79\%) showed flux rope morphologies in XRT data as
well. A characteristic example (event 1 of  Table 1) is presented in
Figure 11 (panels a and b); the appearance of the flux rope is
similar in both images. The three EFRs that were not detected by the 
XRT were rather weak events whose flux rope topology was hard to discern 
probably due to saturation from the flare. 

Table 2 indicates that XRT detected flux rope morphologies in 6 events
that showed no evidence for hot flux ropes in the AIA data (one PE,
three  PFLs, and 2 CFLs). In all 6 events the flux ropes were detected
in thin-filter images only. A typical example appears in panels c and
d of Figure 11 that show 131 \AA\ and XRT observations, respectively,
of event 53.  The AIA image shows a PFL event (eruption without hot
flux rope or  prominence) whereas the XRT image shows a
flux-rope-candidate feature that is indicated by an arrow. Its
morphology bears some resemblance to the hot flux rope of event 89
(Figure 3f). The flux rope of event 53 might be too cool to show in
131 \AA\ and too hot to show in 171 and 304 \AA, but it may have the
right temperature to appear in the broadband  Be thin XRT image; its
temperature may be higher than 2 MK and below 10 MK. Similar arguments
apply to the other five events that showed flux ropes in XRT but not
in AIA.

\section{Conclusions}

This is the first large-scale survey of EUV flare observations to
assess whether hot (10~MK) flux ropes are common, irrespective of
their formation time relative to the CME eruption. We considered only
M- and X-class flare events as hot flux ropes are more likely to exist
in association to large flares. The flux ropes were  identified in 131
\AA\ data using certain morphological criteria (Section~2).  They were
identified as {\it hot} flux ropes if they were  visible only in  131
\AA\ data and not in the 171 and 304 \AA\ data. A summary of our
results is presented in Table~3 and in the pie diagram of Figure
12. The main conclusions of our work are as follows.

1) 32\% of the flares in our database contain clear hot flux rope
morphologies (45/141).  In 34 cases the flux rope erupts (EFR events)
and in 11 cases it does not  (CFR events).  There were 70 flares
associated with CMEs so a hot flux rope configuration was  involved in
49\% of the eruptive events. 

2) The number of confined events with hot flux ropes is about three
times smaller than the number of eruptive  hot flux rope events
whereas half of the flares of our catalog were confined. It is likely
that confined hot flux ropes reach lower altitudes than eruptive ones
thus hindering their identification.

3) Only a small number of hot flux ropes is seen face-on (i.e. with
their axis perpendicular to the line of sight). Most of them are
viewed either  edge-on (i.e. with their axis parallel to the line of
sight) or at  intermediate angles. The smaller number of flux ropes
seen face-on may result from two factors: (1) the natural tendency of
AR neutral lines to lie parallel to the LOS and (2) the weaker signal
due to the small LOS integration path for face-on structures compared
to edge-on ones.

4) The database contains 19 prominence eruptions without signatures of
hot flux rope and 7 cases of eruptive hot flux ropes accompanied with
prominence material. In the latter, the hot ejected plasma  appears
either intermingled with the cooler ejected plasma or spatially
separated from it.

5) 33 of the 34 events with eruptive hot flux ropes were associated
with CMEs and 27 of them (about 80\%) were clear flux rope CMEs. The
remaining 6 might also be flux rope CMEs but the flux rope topology
was hard to discern for several reasons: they may propagate at large angles
from the plane of the sky or through areas disturbed by previous
events or they may be too compact to discern their flux rope
morphology (Vourlidas et al. 2013).  We conclude that a hot flux rope
morphology in the EUV is a very good predictor for a flux rope CME.

Based on {\it Yohkoh} SXT and LASCO data, Kim  et al. (2005b)
found higher percentages of soft X-ray ejecta  associated with
CMEs. However,  their dataset  contained a smaller  percentage of
confined events than ours.

6) Table 3 shows that 60\% of the observed CMEs in our sample have
flux rope structures while Vourlidas et al. (2013) found that at least
40\% of the LASCO CMEs between 1997 and 2010 had flux rope structures.
Our higher percentage could be due to a combination of reasons. First,
most of our events occurred close to the limb which makes the
identification of a flux rope morphology easier. Second, our database
contains large events that presumably are associated with stronger
than average reconnection thus facilitating the formation of hot flux
ropes. Finally, lower percentage of flux rope CMEs is not surprising
given the fact that the Vourlidas et al. (2013) statistics include
both quiet Sun and active region eruptions that may or may not be
associated with flares. 

7) 40 of the events were also observed by XRT on Hinode. 35\% of
these flares contained hot flux ropes in AIA  while 43\% of
them showed flux ropes in XRT data. Therefore the XRT data show
that flux ropes are a rather common ingredient in events
associated with M- and X-class flares. Flux ropes were
detected in half of the 40 flares irrespective of the instrument
used for their detection.

We found that 49\% of our events have hot flux ropes but practically
all CME models expect the erupting structure to be a flux rope. This
raises the question of the magnetic configuration in the remaining
51\% of the database (36 cases). Is there a magnetic flux rope
configuration or not? We believe that there is a straightforward
explanation. The flux rope exists but it is simply too cool to be
detected in 131\AA\ and possibly too hot to show in 171 \AA, 304 \AA, or
even, 193 \AA. There is quite a substantial amount of evidence to
support our assertion: (i) The higher percentage of flux rope CMEs (60\%)
suggests that 20\% of the flux ropes, in our study, were not visible
in 131 \AA. (ii) the strong association between \textit{hot\/} flux
ropes and flux rope CMEs (80\%) reflects the expected association between the
length of the erupting neutral line, the amount of free magnetic
energy available for the eruption and subsequently the percentage of
magnetic energy available to heat the plasma. Since the maximum temperature attained during impulsive heating events is proportional to the volumetric heating (e.g., Cargill 1994; Patsourakos and Klimchuk 2006) it is anticipated that longer (shorter) flux ropes will reach lower (higher) temperatures, for a given magnetic field thus magnetic free energy. We, therefore expect a spectrum of erupting
flux ropes with temperatures correlated to the available magnetic
energy. As further support, we submit (i) the detection of flux ropes
with the XRT in 6 events that did not show such signatures  
in 131-\AA\ data (see Section 6) and (ii) the detection of a flux rope in
284 \AA\ ($\sim 1.8$ MK) but not in 195 \AA\ reported by Vourlidas et al
(2012) for an AR eruption during the early rise of Cycle 24. Finally,
the appearance of hot flux ropes could  be compromised by
projection effects, loops seen along the line of sight  and saturated
emission. This could be especially problematic for face-on cases. 

Based on the above discussion, we conclude that our estimate of 32\%
for the rate of occurrence of hot flux ropes in the flares above M
class, and of 49\% occurrence rate of hot flux ropes in the eruptive
events, constitute lower limits.

An obvious extension of this work is a survey of our database to
assess the timing of the formation of the eruptive hot flux ropes with
respect to  the initiation of the associated CMEs.

\acknowledgments

We thank the referee for his/her constructive comments.
This research has been partly co-financed by the European Union
(European Social Fund -ESF) and Greek national funds through the
Operational Program ``Education and Lifelong Learning" of the National
Strategic Reference Framework (NSRF) -Research Funding Program:
``Thales. Investing in knowledge society through the European Social
Fund". SP acknowledges support from an FP7 Marie Curie Grant
(FP7-PEOPLE-2010-RG/268288). AV was also supported by internal JHU/APL funds.

\clearpage

\begin{figure*}
\epsscale{1.00}
\plotone{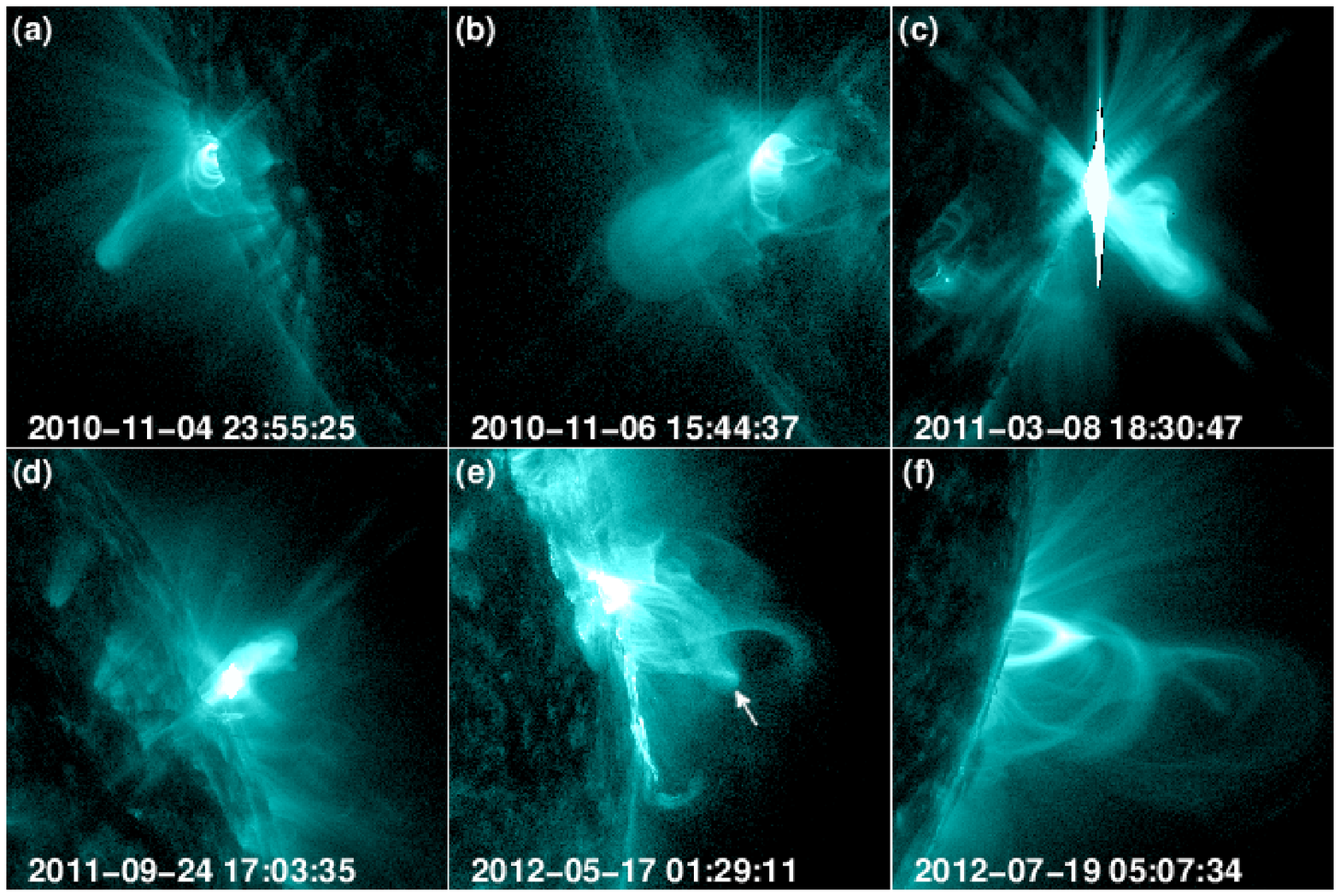}
\caption{131 \AA\ snapshots of various events with hot flux ropes seen 
edge-on. Panels (a) to (f) correspond to events 1, 2, 12, 28, 61, and 69, 
in Table~1, respectively. The arrow marks the hot flux rope of panel (e). Date and time 
appears at the bottom of each image. The field of view in each image is 420 
$\times$ 420 arcsec$^2$. In these and subsequent solar images, solar north is 
up, solar west to the right. (A color version of this figure is available in 
the online journal.)}
\end{figure*}

\begin{figure*}
\epsscale{1.00}
\plotone{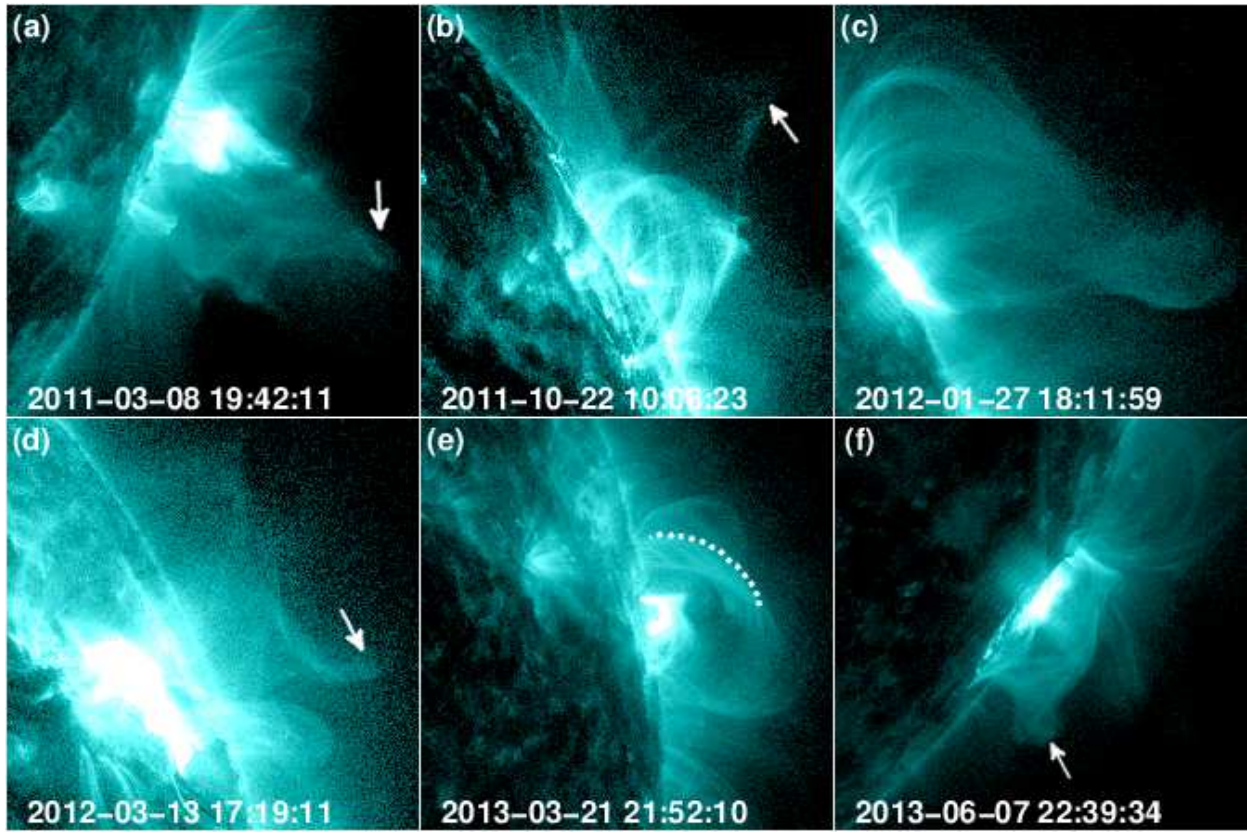}
\caption{131 \AA\ snapshots of various events with hot flux ropes seen
face-on. Panels (a) to (f) correspond to events 13, 35, 49, 54, 87, and 99, 
in Table~1, respectively. When appropriate arrows are used to mark the hot flux ropes.
The dotted curve delineates the outer edge of the hot flux rope of panel (e).
The field of view in each image is 420 $\times$ 420 arcsec$^2$.
(A color version of this figure is available in the online journal.)}
\end{figure*}

\begin{figure*}
\epsscale{1.00}
\plotone{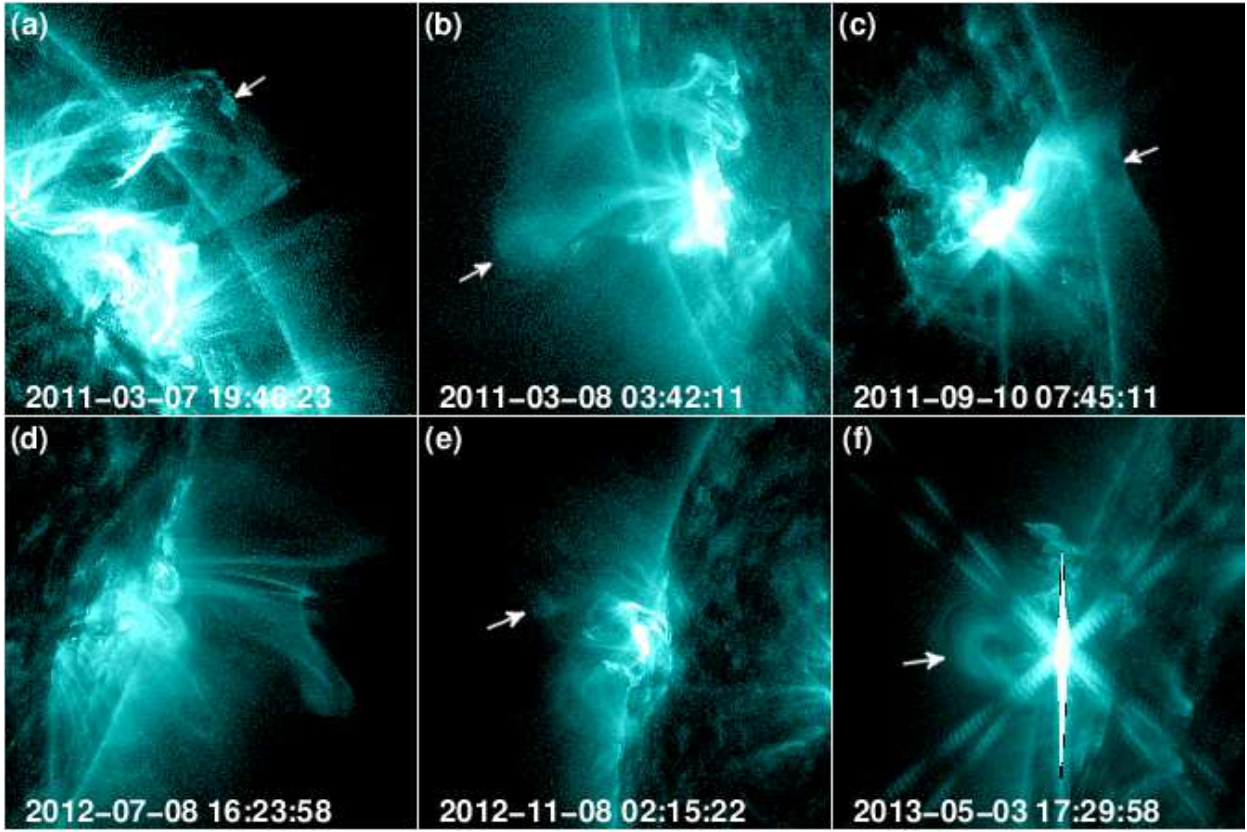}
\caption{131 \AA\ snapshots of various events with hot flux ropes viewed from
intermediate angles. Panels (a) to (f) correspond to events 8, 11, 23, 67, 81, 
and 89, respectively. When appropriate arrows are used to mark the hot flux 
ropes. The field of view in each image is 420 $\times$ 420 arcsec$^2$.
(A color version of this figure is available in the online journal.)}
\end{figure*}

\begin{figure*}
\epsscale{1.00}
\plotone{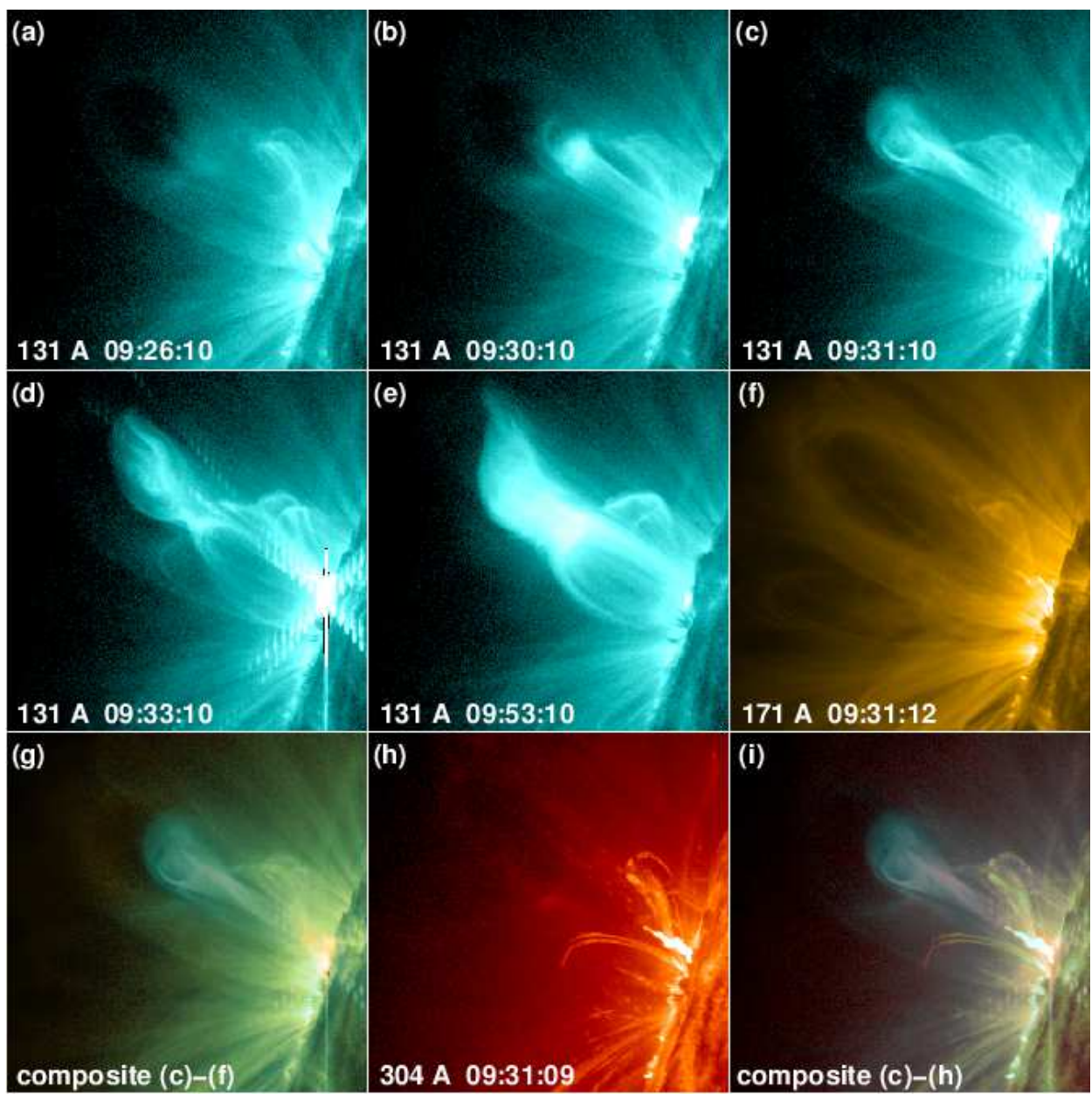}
\caption{Example of a confined event with hot flux rope (CFR). The images
come from event 86 that occurred on 2013 January 5. Panels (a)-(e) show
131 \AA\ images and panels (f) and (h) show 171 and 304 \AA\ images, 
respectively. Panels (g) and (i) are color composites made from the images
of panels (c)-(f) and (c)-(h), respectively. The colors used in the composite
images are those used for the display of the 131, 171, and
304 \AA\ images. The field of view is 240 $\times$ 240 arcsec$^2$.
(An animation (movie1.mp4) and color version of this figure are available in 
the online journal.)}
\end{figure*}

\begin{figure*}
\epsscale{1.00}
\plotone{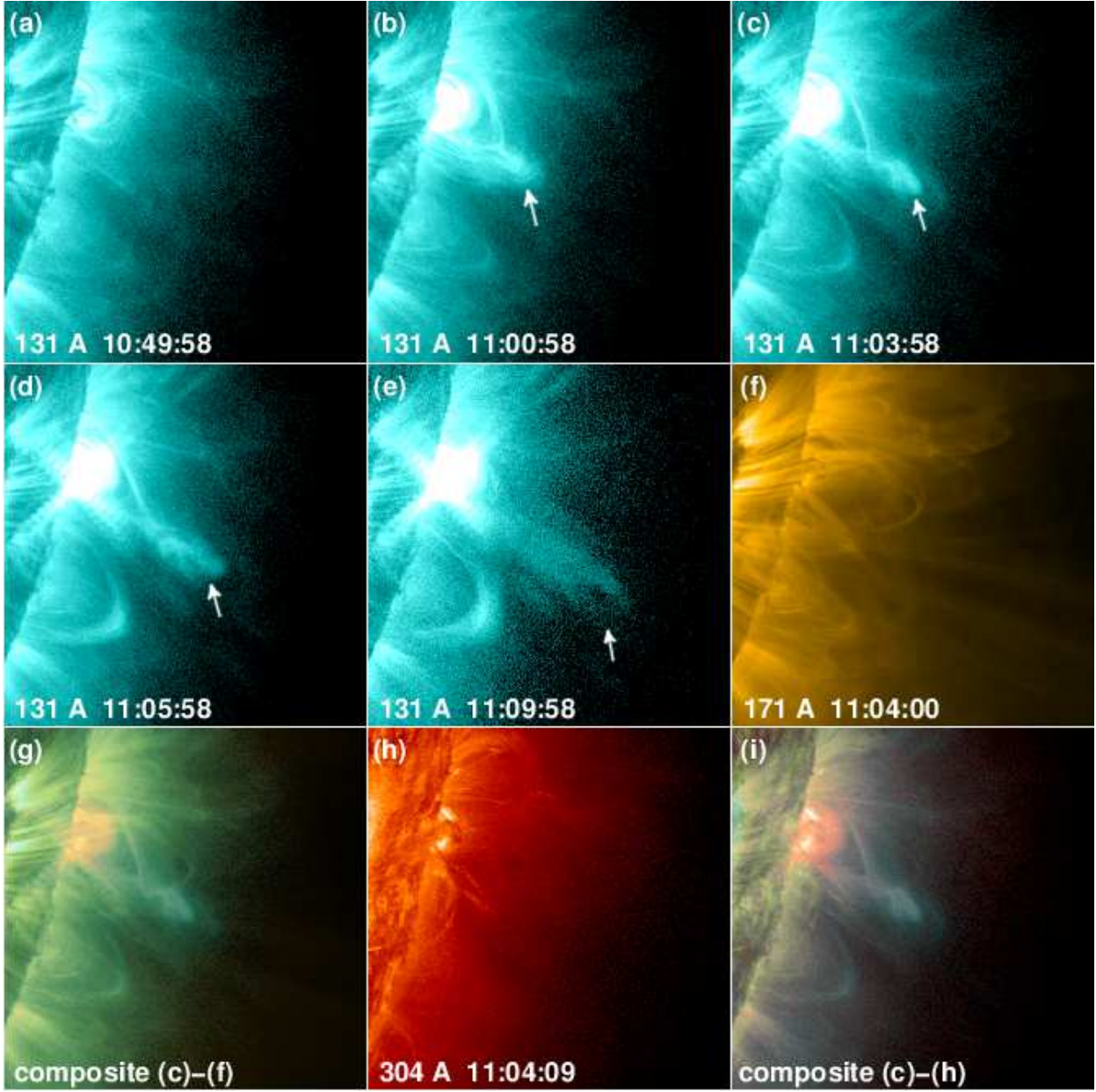}
\caption{Example of an eruptive event with hot flux rope (EFR). The images
come from event 115 that occurred on 2013 November 21. The format of the figure
is the same as the format of Figure 4. Arrows are used to mark the hot flux
rope. The field of view is 300 $\times$ 300 arcsec$^2$. 
(An animation (movie2.mp4) and color version of this figure are available in 
the online journal.)}
\end{figure*}

\begin{figure*}
\epsscale{1.00}
\plotone{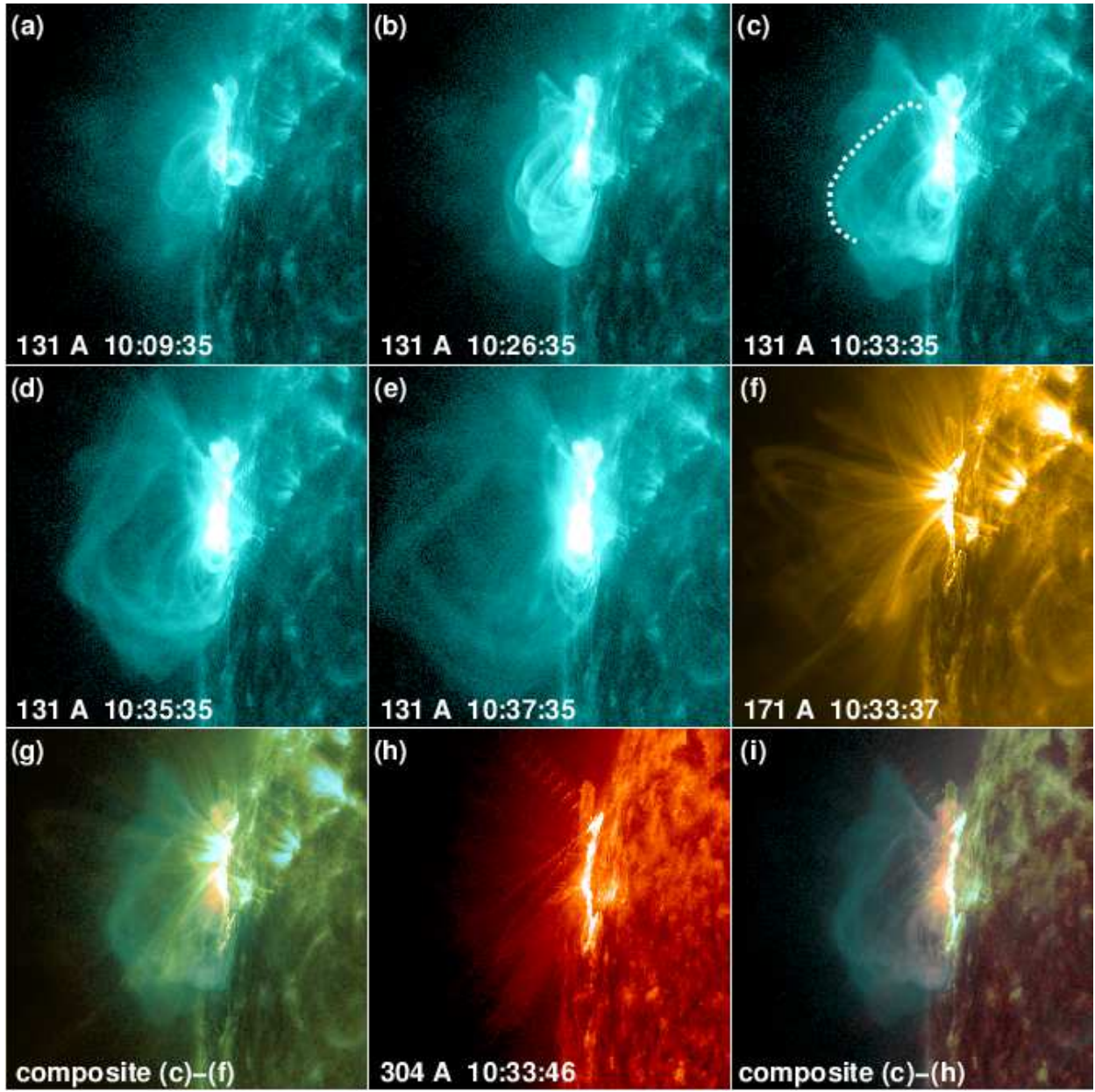}
\caption{Example of an eruptive event with hot flux rope (EFR). The images
come from event 25 that occurred on 2011 September 22. The format of the figure
is the same as the format of Figure 4. The dotted curve of panel (c) delineates
the outer edge of the hot flux rope. The field of view is 480 $\times$ 
480 arcsec$^2$. 
(An animation (movie3.mp4) and color version of this figure are available in 
the online journal.)}
\end{figure*}

\begin{figure*}
\epsscale{1.00}
\plotone{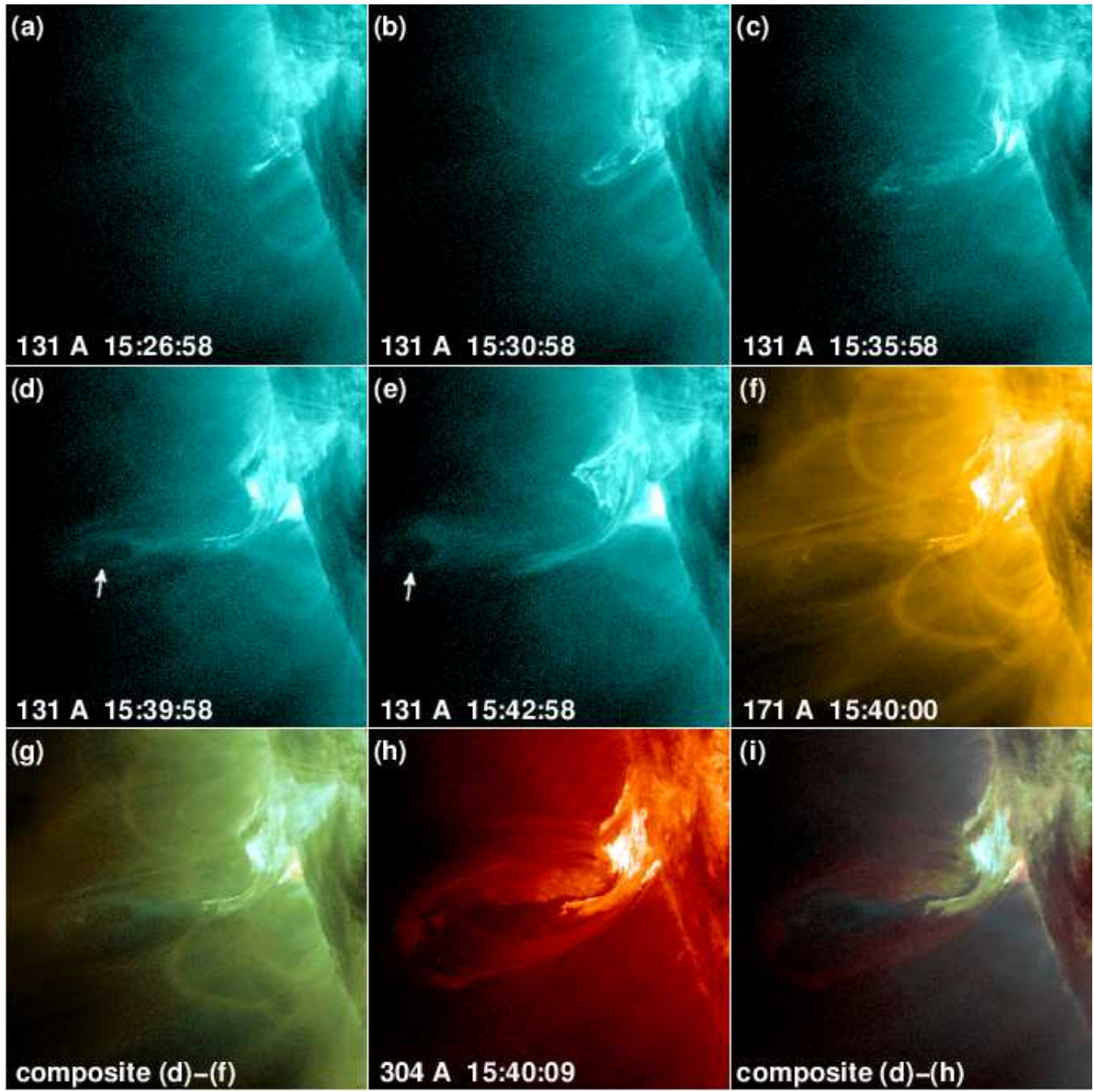}
\caption{Example of an eruptive hot flux rope event (EFR) accompanied with
prominence material. The images come from  event 130 that occurred
on 2014 February 9. The format of the figure is the same as the
format of Figure 4 with the exception that the color composites of
panels (g) and (i) have been made from the images of panels (d)-(f)
and (d)-(h), respectively.  Arrows are used to mark the hot flux
rope. The field of view is 300 $\times$ 300 arcsec$^2$. 
(An animation (movie4.mp4) and color version of this figure are available in 
the online journal.)}
\end{figure*}

\begin{figure*}
\epsscale{1.00}
\plotone{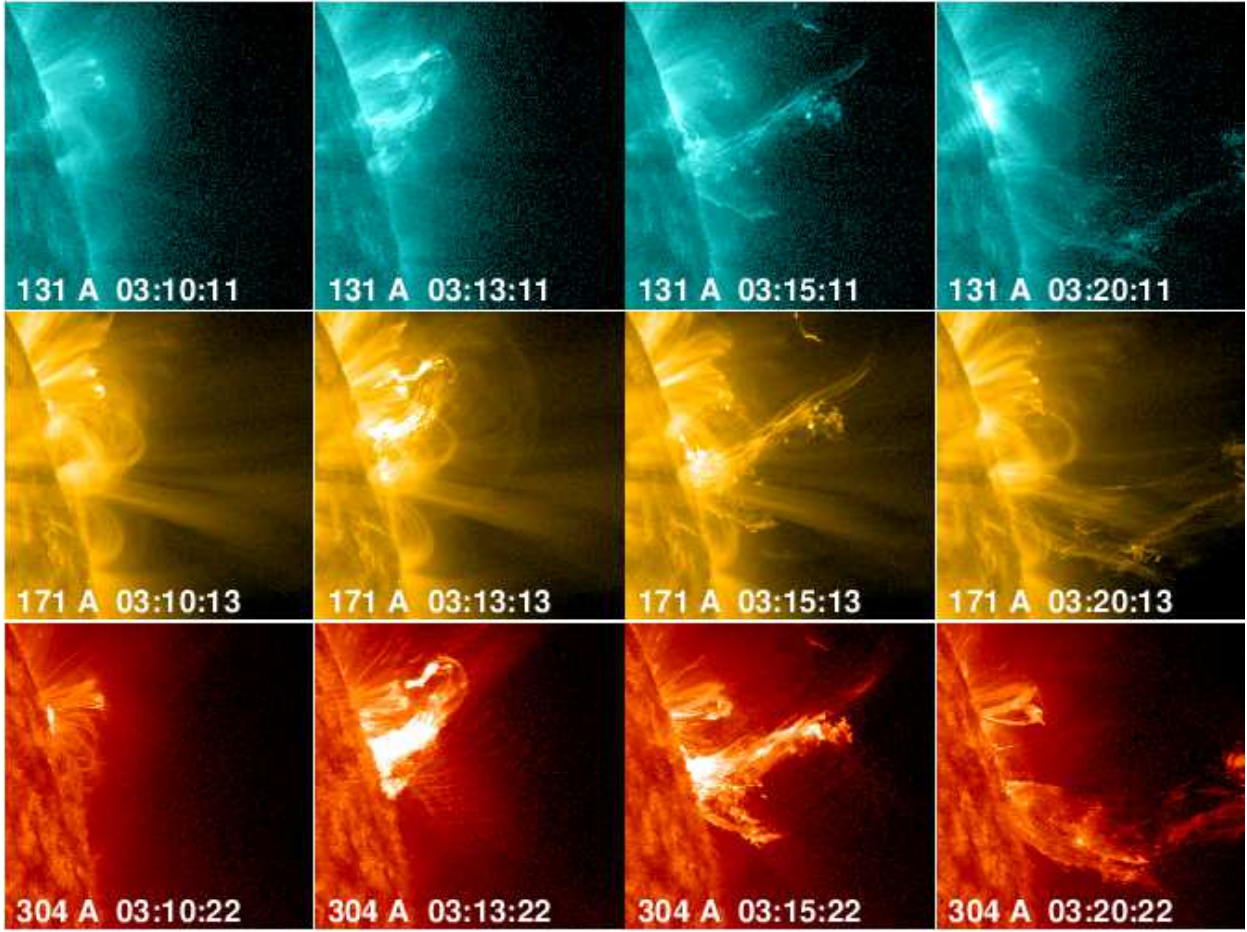}
\caption{Example of a prominence eruption without hot flux rope (PE). Snapshots
from the 131 (top), 171 (middle), and 304 (bottom) passband data of event 33 
that occurred on 2011 October 20. The field of view is 300 $\times$ 300 
arcsec$^2$.
(A color version of this figure is available in the online journal.)}
\end{figure*}

\begin{figure*}
\epsscale{1.00}
\plotone{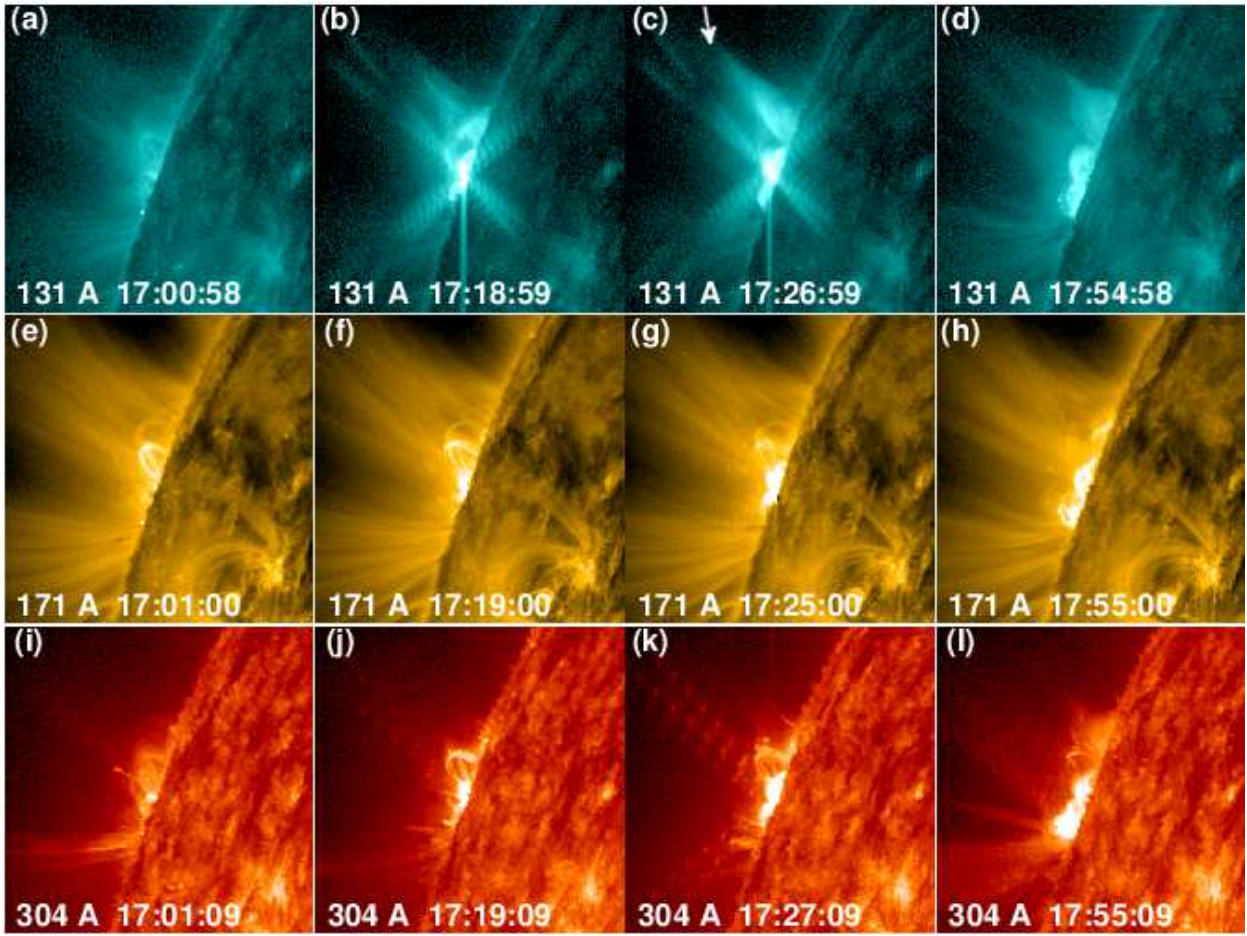}
\caption{Example of an eruptive event without hot flux rope or prominence
(PFL). Snapshots from the 131 (top), 171 (middle), and 304 (bottom) 
passband data of event 72 that occurred on 2012 August 17. The field of view 
is 240 $\times$ 240 arcsec$^2$.
(A color version of this figure is available in the online journal.)}
\end{figure*}

\begin{figure*}
\epsscale{1.00}
\plotone{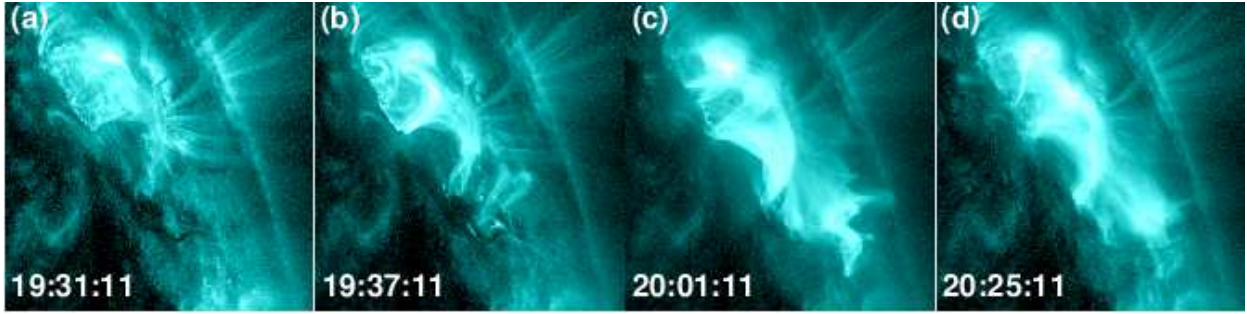}
\caption{Example of a confined flare without hot flux rope
(CFL). Snapshots from  the 131 passband data of event 50 that
occurred on 2012 February 6. The field of view is 240 $\times$ 240 
arcsec$^2$.
(A color version of this figure is available in the online journal.)}
\end{figure*}

\begin{figure*}
\epsscale{0.6}
\plotone{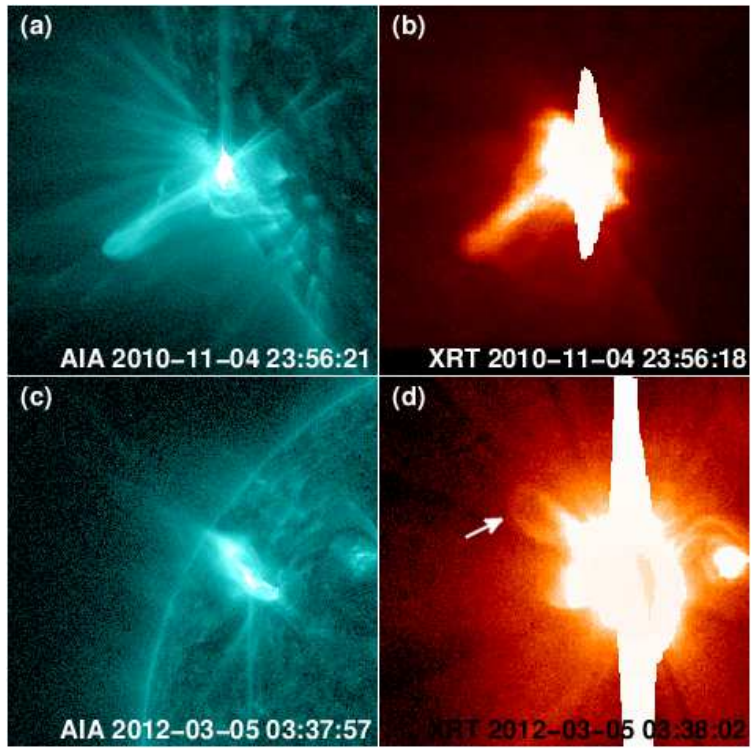}
\caption{131 \AA\ AIA (left column) and XRT (right column) images of events
1 (panels a and b) and 53 (panels c and d). The XRT images of panels (b) and
(d) have been taken with the Al mesh and Be thin filters, respectively. The
field of view of panels (a) and (b) is 400 $\times$ 400 arcsec$^2$ and the 
field of view of panels (c) and (d) is 800 $\times$ 800 arcsec$^2$. The arrow
in panel (d) is used to mark the flux rope. 
(A color version of this figure is available in the online journal.)}
\end{figure*}

\begin{figure}
\epsscale{0.5}
\plotone{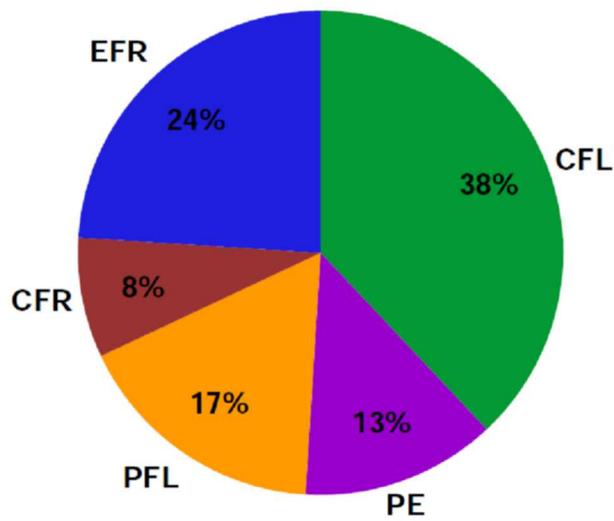}
\caption{Pie diagram that shows the percentages of occurrence of the
different classification groups for hot flux ropes. The acronyms are explained in Section 3.
(A color version of this figure is available in the online journal.)}
\end{figure}

\clearpage

\begin{deluxetable}{lrrrrrrr}
\tablewidth{0pt}
\tablecaption{Catalog of events and their classification according to AIA data}
\tablehead{
\colhead{\#}      & \colhead{Date}      &
\colhead{Flare Start}  &   \colhead{Location} & \colhead{AR}     & \colhead{GOES}  &
\colhead{CME}          & \colhead{Classification\tablenotemark{a}} \\
\colhead{} & \colhead{} & 
\colhead{\& Peak}  &   \colhead{} & \colhead{}     & \colhead{class}  &
\colhead{}          & \colhead{}
}
\startdata
1        & 04 Nov 2010 & 23:30 23:58  & S20E76 & 11121 & M1.6 & No  & CFR \\
2        & 06 Nov 2010 & 15:27 15:36  & S19E58 & 11121 & M5.4 & No  & CFR \\
3        & 28 Jan 2011 & 00:44 01:03  & N16W88 & 11149 & M1.3 & Yes (FR\tablenotemark{b}) & PFL  \\
4        & 09 Feb 2011 & 01:23 01:31  & N16W70 & 11153 & M1.9 & No  &   PFL \\
5        & 24 Feb 2011 & 07:23 07:35  & N14E87 & 11163 & M3.5 & Yes (FR) & EFR \\
6        & 07 Mar 2011 & 07:49 07:54  & S18W75 & 11165 & M1.6 &  No & PFL \\
7        & 07 Mar 2011 & 09:14 09:15  & S17W77 & 11165 & M1.8 &  No & CFL \\
8        & 07 Mar 2011 & 19:43 20:12  & N30W50 & 11164 & M3.7 & Yes (FR) & EFR \\
9        & 07 Mar 2011 & 21:45 21:50  & S17W82 & 11165 & M1.5 & No  & CFL \\
10       & 08 Mar 2011 & 02:24 02:29  & S17W80 & 11165 & M1.3 & No  & CFL \\
11       & 08 Mar 2011 & 03:37 03:58  & S21E72 & 11171 & M1.5 & Yes (FR) & EFR  \\
12       & 08 Mar 2011 & 18:08 18:28  & S17W88 & 11165 & M4.4 & No  & CFR  \\
13       & 08 Mar 2011 & 19:35 20:16  & S19W87 & 11165 & M1.5 & Yes (FR) & EFR \\
14       & 29 May 2011 & 10:08 10:33  & S20E64 & 11226 & M1.4 & Yes (FR) & PFL \\
15       & 07 Jun 2011 & 06:16 06:30  & S22W53 & 11226 & M2.5 & Yes (FR) & PE \\
16       & 14 Jun 2011 & 21:36 21:47  & N14E78 & 11236 & M1.3 & No  & CFR \\
17       & 08 Aug 2011 & 18:00 18:10  & N15W62 & 11263 & M3.5 & No  & CFL \\
18       & 09 Aug 2011 & 03:19 03:45  & N17W69 & 11263 & M2.5 & No  & CFL \\
19       & 09 Aug 2011 & 07:48 08:05  & N14W69 & 11263 & X6.9 & Yes (without FR) & PE \\
20       & 04 Sep 2011 & 11:21 11:45  & N18W84 & 11274 & M3.2 & No  & CFL  \\
21       & 05 Sep 2011 & 04:08 04:28  & N18W87 & 11286 & M1.6 & No  & CFL \\
22       & 05 Sep 2011 & 07:27 07:58  & N18W87 & 11286 & M1.2 & No  & CFL \\
23       & 10 Sep 2011 & 07:18 07:40  & N14W64 & 11283 & M1.1 & Yes (without FR)  & EFR \\
24       & 21 Sep 2011 & 12:04 12:23  & N15E88 & 11301 & M1.8 & No  & CFL \\
25       & 22 Sep 2011 & 10:29 11:01  & N09E89 & 11302 & X1.4 & Yes (FR) & EFR \\
26       & 23 Sep 2011 & 01:47 01:59  & N24W64 & 11300 & M1.6 & No  & CFL \\
27       & 24 Sep 2011 & 09:21 09:40  & N13E61 & 11302 & X1.9 & No  & CFL \\
28       & 24 Sep 2011 & 16:35 16:59  & N23W87 & 11289 & M1.8 & No  & CFR \\
29       & 24 Sep 2011 & 21:23 21:27  & S29W67 & 11303 & M1.2 & No  & CFL \\
30       & 24 Sep 2011 & 23:45 23:57  & S28W66 & 11303 & M1.0 & Yes (without FR) & PFL \\
31       & 25 Sep 2011 & 02:27 02:31  & N22W87 & 11289 & M4.4 & No  & CFL \\
32       & 25 Sep 2011 & 09:25 09:35  & S28W71 & 11303 & M1.5 & Yes (without FR) & PFL \\
33       & 20 Oct 2011 & 03:10 03:25  & N18W88 & 11312 & M1.6 & Yes (without FR) &  PE \\
34       & 21 Oct 2011 & 12:53 13:00  & N05W79 & 11319 & M1.3 & Yes (without FR) &  PE \\
35       & 22 Oct 2011 & 09:18 11:10  & N27W87 & 11314 & M1.3 & Yes (FR) & EFR \\
36       & 31 Oct 2011 & 14:55 15:07  & N20E88 & 11337 & M1.1 & No   & CFL \\
37       & 31 Oct 2011 & 17:21 18:07  & N21E88 & 11337 & M1.4 & No   & PFL \\
38       & 31 Oct 2011 & 18:29 18:33  & N20E88 & 11337 & M1.2 & No   & CFL \\
39       & 02 Nov 2011 & 21:52 22:01  & N20E77 & 11339 & M4.3 & No   & CFL \\
40       & 03 Nov 2011 & 10:58 11:11  & N20E70 & 11339 & M2.5 & No   & CFL \\
41       & 03 Nov 2011 & 20:16 20:27  & N21E64 & 11339 & X1.9 & Yes (without FR) & PFL \\
42       & 03 Nov 2011 & 23:28 23:31  & N20E62 & 11339 & M2.1 & No   & CFL \\
43       & 15 Nov 2011 & 09:03 09:12  & N21W72 & 11348 & M1.2 & No   & CFL \\
44       & 15 Nov 2011 & 22:27 22:31  & N18W81 & 11339 & M1.1 & No   & PE \\
45       & 29 Dec 2011 & 13:40 13:50  & S25E70 & 11389 & M1.9 & No   & CFL \\
46       & 29 Dec 2011 & 21:43 21:51  & S25E67 & 11389 & M2.0 & No   & CFL \\
47       & 30 Dec 2011 & 03:03 03:09  & S25E65 & 11389 & M1.2 & No   & CFL \\
48       & 14 Jan 2012 & 13:14 13:18  & N14E88 & 11401 & M1.4 & No   & CFR \\
49       & 27 Jan 2012 & 17:37 18:36  & N33W85 & 11402 & X1.7 & Yes (FR) &  EFR \\
50       & 06 Feb 2012 & 19:31 20:00  & N19W62 & 11410 & M1.0 & No   & CFL \\
51       & 02 Mar 2012 & 17:29 17:46  & N18E87 & 11429 & M3.3 & Yes (FR) & PFL \\
52       & 04 Mar 2012 & 10:29 10:52  & N16E65 & 11429 & M2.0 & Yes (without FR) & EFR \\
53       & 05 Mar 2012 & 02:30 04:05  & N19E58 & 11429 & X1.1 & Yes (FR) & PFL \\
54       & 13 Mar 2012 & 17:12 17:30  & N17W66 & 11429 & M7.9 & Yes (FR) & EFR \\
55       & 23 Mar 2012 & 19:34 19:40  & S23E87 & 11445 & M1.0 & No  & CFR \\
56       & 16 Apr 2012 & 17:24 17:40  & N14E88 & 11461 & M1.7 & Yes (without FR) & PE \\
57       & 05 May 2012 & 13:19 13:23  & N11E78 & 11476 & M1.4 & No LASCO data & PE  \\
58       & 05 May 2012 & 22:56 23:01  & N11E73 & 11476 & M1.3 & No LASCO data  & CFL \\
59       & 06 May 2012 & 01:12 01:18  & N11E73 & 11476 & M1.2 & No LASCO data  & CFL \\
60       & 06 May 2012 & 17:41 17:47  & N11E63 & 11476 & M1.3 & No LASCO data  & CFL \\
61       & 17 May 2012 & 01:25 01:47  & N07W88 & 11476 & M5.1 & Yes (FR) & EFR \\
62       & 06 Jul 2012 & 13:26 13:30  & S17E85 & 11519 & M1.2 & No  & CFL \\
63       & 07 Jul 2012 & 08:18 08:28  & S16E76 & 11520 & M1.0 & No  & CFL \\
64       & 08 Jul 2012 & 05:41 05:46  & S16W70 & 11515 & M1.3 & No  & CFL \\
65       & 08 Jul 2012 & 09:44 09:53  & S16W70 & 11515 & M1.1 & Yes (FR) & EFR \\
66       & 08 Jul 2012 & 12:06 12:10  & S16W72 & 11515 & M1.4 & No  & CFL \\
67       & 08 Jul 2012 & 16:23 16:32  & S14W86 & 11515 & M6.9 & Yes (without FR) & EFR \\
68       & 17 Jul 2012 & 12:03 17:15  & S15W88 & 11520 & M1.7 & Yes (without FR) & PFL \\
69       & 19 Jul 2012 & 04:17 05:58  & S13W88 & 11520 & M7.7 & Yes (FR) & EFR \\
70       & 27 Jul 2012 & 17:17 17:26  & S24E71 & 11532 & M2.7 & Yes (without FR) & EFR \\
71       & 17 Aug 2012 & 13:12 13:19  & N18E88 & 11548 & M2.4 & Yes (without FR) & PFL \\
72       & 17 Aug 2012 & 17:08 17:20  & N19E87 & 11548 & M1.0 & Yes (without FR) & PFL \\
73       & 18 Aug 2012 & 03:17 03:23  & N19E87 & 11548 & M1.8 & Yes (without FR) & PE \\
74       & 18 Aug 2012 & 16:02 16:09  & N19E80 & 11548 & M2.0 & Yes (without FR) & PE \\
75       & 18 Aug 2012 & 22:46 22:54  & N19E78 & 11548 & M1.0 & No & CFR \\
76       & 18 Aug 2012 & 23:15 23:22  & N21E76 & 11548 & M1.3 & Yes (FR) & PE \\
77       & 30 Aug 2012 & 12:02 12:11  & S27E85 & 11563 & M1.3 &  No & CFR \\
78       & 30 Sep 2012 & 04:27 04:33  & N12W81 & 11583 & M1.3 &  No & CFL \\
79       & 09 Oct 2012 & 23:22 23:27  & S29E86 & 11590 & M1.7 &  No & CFL \\
80       & 10 Oct 2012 & 04:51 05:04  & S29E86 & 11590 & M1.0 &  No & CFL \\
81       & 08 Nov 2012 & 02:08 02:23  & N13E89 & 11611 & M1.7 &  Yes (FR) & EFR \\
82       & 11 Nov 2012 & 02:11 02:28  & N15E89 & 11614 & M1.0 &  No & CFR \\
83       & 20 Nov 2012 & 12:36 12:41  & N11W89 &       & M1.7 &  Yes (without FR) & PE  \\
84       & 27 Nov 2012 & 15:52 15:57  & N06W72 & 11618 & M1.6 &  No & CFL \\
85       & 28 Nov 2012 & 21:20 21:28  & S12W58 & 11620 & M2.2 &  No & CFL \\
86       & 05 Jan 2013 & 09:26 09:31  & N20E88 & 11652 & M1.7 &  No & CFR \\
87       & 21 Mar 2013 & 21:42 22:04  & N09W88 & 11692 & M1.6 &  Yes (FR) & EFR \\
88      & 05 Apr 2013 & 17:34 17:48  & N07E88 & 11719 & M2.2 &  No & CFL \\
89      & 03 May 2013 & 17:24 17:32  & N15E83 & 11739 & M5.7 & Yes (FR) & EFR  \\
90      & 12 May 2013 & 20:17 20:31  & N10E89 & 11748 & M1.9 & Yes (FR)  & PFL \\
91      & 12 May 2013 & 22:37 22:44  & N10E89 & 11748 & M1.2 &  No & CFL \\
92      & 13 May 2013 & 01:53 02:16  & N11E89 & 11748 & X1.7 & Yes (FR) & PFL \\
93      & 13 May 2013 & 11:57 12:03  & N10E89 & 11748 & M1.3 & No  & CFL \\
94      & 13 May 2013 & 15:48 16:05  & N08E89 & 11748 & X2.8 & Yes (FR)  & EFR \\
95      & 13 May 2013 & 23:59 01:11\tablenotemark{c}  & N08E77 & 11748 & X3.2 & Yes (FR)  & EFR \\
96      & 15 May 2013 & 01:25 01:48  & N10E68 & 11748 & X1.2 & Yes (FR) & EFR \\
97      & 20 May 2013 & 04:45 05:25  & N09E89 & 11753 & M1.8 & Yes (FR) & EFR \\
98      & 22 May 2013 & 13:08 13:32  & N14W87 & 11745 & M5.0 & Yes (FR) &  EFR \\
99      & 07 Jun 2013 & 22:11 22:49  & S32W89 & 11762 & M5.9 & Yes (FR) & EFR \\
100     & 21 Jun 2013 & 02:30 03:00  & S14E73 & 11777 & M2.9 & Yes (FR) & EFR \\
101     & 03 Jul 2013 & 07:00 07:08  & S14E82 & 11785 & M1.5 & Yes (FR) & PFL \\
102     & 09 Oct 2013 & 01:23 01:48  & S23E71 & 11865 & M2.8 & Yes (without FR) & PE \\
103     & 25 Oct 2013 & 02:48 03:02  & S07E76 & 11882 & M2.9 & Yes (without FR) & EFR \\
104     & 25 Oct 2013 & 07:53 08:01  & S08E73 & 11882 & X1.7 & Yes (FR) & PFL \\
105     & 25 Oct 2013 & 09:43 10:11  & S08E73 & 11882 & M1.0 & No  & CFL \\
106     & 25 Oct 2013 & 14:51 15:03  & S06E69 & 11882 & X2.1 & Yes (FR) & PFL \\
107     & 26 Oct 2013 & 19:22 19:27  & S12E87 & 11884 & M3.1 & Yes (FR)  & EFR \\
108     & 27 Oct 2013 & 12:36 12:48  & S11E73 & 11884 & M3.5 & No & EFR \\
109     & 28 Oct 2013 & 01:41 02:03  & N05W72 & 11875 & X1.0 & Yes (FR) & EFR \\
110     & 28 Oct 2013 & 04:32 04:41  & N08W72 & 11875 & M5.1 & Yes (without FR) & PE \\
111     & 28 Oct 2013 & 14:00 14:05  & N08W78 & 11875 & M2.8 & Yes (without FR) & PFL \\
112     & 06 Nov 2013 & 23:35 00:02\tablenotemark{c}  & S11W88 & 11882 & M1.9 & Yes (FR)  & PE \\
113     & 11 Nov 2013 & 11:01 11:18  & S17E74 & 11897 & M2.4 & No & PFL \\
114     & 19 Nov 2013 & 10:14 10:26  & S13W69 & 11893 & X1.0 & Yes (FR) & PFL \\
115     & 21 Nov 2013 & 10:52 11:11  & S14W89 & 11895 & M1.2 & Yes (FR) & EFR \\
116     & 02 Jan 2014 & 02:24 02:33  & S05E89 & 11944 & M1.7 & No & CFL \\
117     & 04 Jan 2014 & 22:09 22:52  & S14W89 & 11936 & M2.0 & Yes (FR) & EFR \\
118     & 08 Jan 2014 & 03:39 03:47  & N11W88 & 11947 & M3.6 & Yes (FR) & EFR \\
119     & 13 Jan 2014 & 21:48 21:51  & S08W75 & 11944 & M1.3 & Yes (without FR) & PFL \\
120     & 27 Jan 2014 & 01:05 01:22  & S16E88 & 11967 & M1.0 & No & CFL \\
121     & 27 Jan 2014 & 02:02 02:10  & S13E88 & 11967 & M1.1 & No & CFL \\
122     & 27 Jan 2014 & 22:05 22:10  & S14E88 & 11967 & M4.9 & No & CFL \\
123     & 28 Jan 2014 & 04:02 04:09  & S14E88 & 11967 & M1.5 & No & CFL \\
124     & 28 Jan 2014 & 07:25 07:31  & S10E75 & 11967 & M3.6 & Yes (without FR) & PE \\
125     & 28 Jan 2014 & 11:34 11:38  & S10E72 & 11967 & M1.4 & Yes (without FR) & PE \\
126     & 28 Jan 2014 & 12:38 12:46  & S15E79 & 11967 & M1.3 & No & CFL \\
127     & 28 Jan 2014 & 15:24 15:26  & S13E88 & 11967 & M3.5 & No & CFL \\
128     & 28 Jan 2014 & 19:00 19:40  & S14E75 & 11967 & M4.9 & No & CFL \\
129     & 28 Jan 2014 & 22:04 22:16  & S14E74 & 11967 & M2.6 & Yes (without FR) &  PFL \\
130     & 09 Feb 2014 & 15:40 16:17  & S16E88 &       & M1.0 & Yes (FR) & EFR \\
131     & 20 Feb 2014 & 07:26 07:56  & S15W75 & 11976 & M3.0 & Yes (FR) & PE \\
132     & 23 Feb 2014 & 05:50 06:10  & S16E88 & 11990 & M1.1 & Yes (without FR) & PFL \\
133     & 24 Feb 2014 & 11:03 11:17  & S11E88 & 11990 & M1.2 & Yes (without FR) & EFR \\
134     & 25 Feb 2014 & 00:39 00:49  & S12E77 & 11990 & X4.9 & Yes (FR) & PE \\
135     & 01 Mar 2014 & 13:18 13:33  & S12W88 & 11982 & M1.1 & No & CFL \\
136     & 12 Mar 2014 & 10:55 11:05  & N14W70 & 11996 & M2.5 & No & CFL \\
137     & 12 Mar 2014 & 22:28 22:34  & N14W76 & 11996 & M9.3 & No  & CFL \\
138     & 13 Mar 2014 & 19:03 19:19  & N15W87 & 11996 & M1.2 & No & CFL  \\
139     & 20 Mar 2014 & 03:42 03:56  & S12E76 & 12014 & M1.7 & No & CFL \\
140     & 22 Mar 2014 & 06:58 07:02  & S09W69 & 12011 & M1.1 & No & CFL \\
141     & 31 Mar 2014 & 07:20 08:07  & S13W76 & 12014 & M1.4 & Yes (without FR) & PE \\
\enddata
\tablenotetext{a}{The acronyms CFL, PE, PFL, CFR, and EFR are explained in Section 3.}
\tablenotetext{b}{Flux rope.}
\tablenotetext{c}{Next day.}
\end{deluxetable}

\clearpage

\begin{deluxetable}{ccc}
\tablewidth{0pt}
\tablecaption{Events observed by both AIA and XRT}
\tablehead{
\colhead{AIA Classification\tablenotemark{a}}  & \colhead{131 \AA\ AIA}    & \colhead{XRT FRs\tablenotemark{b}} \\
\colhead{} & \colhead{Number of Events} & \colhead{Number of Events}
}
\startdata
CFR &  4 & 4  \\
EFR &  10 & 7  \\
PE  &  3 & 1  \\
PFL &  8 & 3  \\
CFL &  15 & 2   \\
Total &  40 & 17
\enddata
\tablenotetext{a}{The acronyms CFL, PE, PFL, CFR, and EFR are explained in Section 3.}
\tablenotetext{b}{Flux ropes.}
\end{deluxetable}

\clearpage

\begin{deluxetable}{cccc}
\tablewidth{0pt}
\tablecaption{Summary of results}
\tablehead{
\multicolumn{2}{c}{Inner Corona} & \multicolumn{2}{c}{Outer Corona} \\
\colhead{Classification\tablenotemark{a}}  & \colhead{Number of Events}    & \colhead{Number of CMEs} & \colhead{Number of FR\tablenotemark{b} CMEs}}
\startdata
CFR & 11 & 0   & 0 \\
EFR & 34 & 33  & 27 \\
PE  & 19 & 17  & 5 \\
PFL & 24 & 20  & 10 \\
CFL & 53 & 0   & 0 \\
Total & 141 & 70 & 42
\enddata
\tablenotetext{a}{The acronyms CFL, PE, PFL, CFR, and EFR are explained in Section 3.}
\tablenotetext{b}{Flux rope.}
\end{deluxetable}




\end{document}